\documentclass[12pt]{article}

\usepackage{scicite}
\usepackage{times}
\usepackage{cite}
\usepackage{amsthm, amssymb}
\usepackage{comment}
\theoremstyle{definition}

\usepackage{graphicx}
\graphicspath{ {./images/} }
\usepackage{amsmath}
\usepackage{caption}
\usepackage{subcaption}
\usepackage{url}
\usepackage{hyperref}
\usepackage{cleveref}
\usepackage{amsmath}
\usepackage{booktabs}
\Crefname{equation}{Eq.}{Equations}
\crefformat{figure}{Fig.~#2#1#3}

\newenvironment{sciabstract}{%
\begin{quote} \bf}
{\end{quote}}

\title{PREPARE: PREdicting PAndemic's REcurring Waves Amidst Mutations, Vaccination, and Lockdowns}

\author{Narges~M.~Shahtori,$^{1}$, S.~Farokh~Atashzar$^{1,2,3,4\ast}$  \\
\\
\normalsize{$^{1}$Department of Electrical and Computer Engineering, New York University}\\
\normalsize{$^{2}$Department of Mechanical and Aerospace Engineering, New York University}\\
\normalsize{$^{3}$Department of Biomedical Engineering, New York University}\\
\normalsize{$^{4}$NYU Center for Urban Science and Progress (CUSP), New York University}\\
\normalsize{$^\ast$To whom correspondence should be addressed; E-mail:  sfa7@nyu.edu.}
}

\date{}

\begin{document} 


\baselineskip24pt


\maketitle

\begin{sciabstract}

\end{sciabstract}
\section{Teaser}

This study releases an adaptable framework that can provide insights to policymakers to predict the complex recurring waves of the pandemic in the medium postemergence of the virus spread, a phase marked by rapidly changing factors like virus mutations, lockdowns, and vaccinations, offering a way to forecast infection trends and stay ahead of future outbreaks even amidst uncertainty. The proposed model is validated on data from COVID-19 spread in Germany.

\section{Introduction}\label{sec:introduction}
Despite rapid advances in vaccine technology and the progress of medical sciences, infectious diseases remain an imminent danger to human and animal health. Throughout the outbreak, reliable forecasts are critical for policy-makers and decision-makers to implement mitigation strategies based on available resources. Mathematical modeling of infectious disease provides a powerful framework to deploy an informative crisis strategy-based approach \cite{heesterbeek2015modeling}. 
Infectious disease models allow for establishing critical epidemiological parameters such as reproductive ratio, explaining the complex dynamics of infectious diseases, and impacts of control measures\cite{wormser2008modeling}\cite{eubank2004modelling}. However, sparse data, imperfect knowledge about contributing factors, and systematic errors impose several challenges for extrapolating forecasts throughout the medium-term post-emergence of the epidemic.  

While the specific central epidemiological factors can differ from one disease to another, they can be generalized into two categories. Firstly, the inherent properties of the virus play a pivotal role, encompassing factors such as transmission probability, infectious period, virulence, potential for asymptomatic transmission, and the dynamics of immunity and reinfection. Secondly, exogenous variables, including human behaviors, movement patterns, average contact rates, and the effectiveness of pharmaceutical and non-pharmaceutical mitigation strategies. Additionally, environmental factors, socioeconomic conditions, and access to healthcare contribute substantially to disease transmission dynamics\cite{roberts2023quantifying}. One major challenge in modeling infectious diseases, given the aforementioned epidemiological factors, is understanding how these factors interact and influence each other. For example, changes in one factor, such as human mobility, can impact others, such as the effectiveness of quarantine measures~\cite{viboud2018rapidd}\cite{middleton2024modeling}. This, in turn, affects the overall endemic state of the disease, creating a dynamic and interdependent system.

Furthermore, inherent delays associated with implementing mitigation strategies while the outbreak is evolving add to the complexity \cite{dehning2020inferring}. Specifically, delays and partial data may create challenges for developing a reliable system identification framework, leading to the dismissal of essential interactions and uncertainty in estimating model parameters \cite{o2002tutorial}\cite{althouse2015enhancing}\cite{han2019confronting}. The availability and quality of data concerning the affected population, such as the number of infectious individuals and mortality rates, are essential for creating a curated model to accurately forecast the spread and depict an accurate picture of risk associated with mitigation strategies for ongoing or emerging infectious diseases. However, data collection during outbreaks faces significant challenges, mainly due to limited resources and conflicting priorities between disease control and information gathering. These limitations often result in a lack of reliable data, amplifying the modeling uncertainty. 


Conventional differential infectious disease models, such as \textit{susceptible-exposed-infected-recovered} ($SEIR$), are crucial tools for short-term disease spreading forecasting and long-term trend analysis~\cite{metcalf2017opportunities}. However, their effectiveness relies heavily on accurate data regarding key epidemiological factors and reported cases. The $SEIR$ model, in particular, shows significant limitations when applied to medium-term post-emergence scenarios where there is sufficient data to derive central epidemiological parameters, but the variables are quickly changing. Specifically, during this phase, sufficient data regarding epidemiological parameters and the virus’s inherent properties from the first generation of the variant is available. Unlike the initial phase of a disease outbreak, an established reporting system provides data on infected cases and mortality during this phase. Yet reported cases are susceptible to human errors, noise, and delays (such as weekend reporting lags), but the necessary data to model disease progression is available. Due to the inherent characteristics of this phase, stochastic fluctuations in infected cases are inevitable because of the changes in both pathogen virulence and human interaction dynamics. These statistical models often fail to dynamically update these contributing factors, accurately predict the number of infectious, and, ultimately, inadequately capture the complexities of medium-term post-emergence disease dynamics. 

Integrating prior knowledge to estimate critical epidemiological parameters can significantly enhance the prediction and extrapolation of infectious data. Over the last decade, researchers have utilized various methods to enhance understanding of disease dynamics and improve public health responses. One such approach is the augmentation of traditional epidemiological data with other data sources, such as social media data and mobile phone records, to gain real-time insights about disease dissemination, also known as transmission probability locally and globally\cite{lu2019improved}\cite{peixoto2020modeling}\cite{gao2020association}. Additionally, advanced statistical methods, such as Bayesian modeling\cite{mccabe2023alternative} \cite{gozzi2023estimating}\cite{dodd2023transmission} \cite{momenyan2022modeling} and particle filters \cite{calvetti2020bayesian}\cite{shahtori2016sequential}\cite{dutra2020monitoring} are utilized to infer the model's parameters and subsequently forecast the infected cases. Bayesian estimation is particularly well-suited to augment prior knowledge and infectious disease equations to assess parameter uncertainties quantitatively and compensate for imperfect knowledge about hidden states of the system. 

In this study, we propose a hierarchical Bayesian online changing-point detection framework to address the challenges of modeling disease spread during the medium-term post-emergence phase of a pandemic. This framework is designed to enhance the predictive accuracy of infectious disease spread by incorporating genetic algorithms and sliding window mechanisms. The proposed model allows 1) capturing the temporal variation in transmission probability influenced by factors such as virus mutations, developed immunity resilience, and changes in average contact rate due to non-pharmaceutical interventions, 2) Optimizing the selection of the model's initial parameters using genetic algorithm, ensuring the model starts with the most appropriate parameters, minimizing the risk of divergence and, 3) mitigating the impact of partially identified model and noisy data by focusing on smaller subsets of the dataset using sliding window technique, enhancing the model's robustness and predictive power. This approach ensures that the model remains responsive to evolving epidemiological landscapes and updates the prediction of disease progression based on the available data, enhancing its predictive accuracy and utility in public health interventions.

We applied the model to data collected in Germany, targeting data from September 2020 to January 2021, which exhibits the medium-term post-emergence features. This allows us to shed light on the significance of new variants of SARS-CoV-2 in Germany with respect to the implementation of pharmaceutical and non-pharmaceutical mitigation strategies. We utilized the extensive dataset provided by Johns Hopkins University's COVID-19 repository \cite{dong2020covid} as a cornerstone for our analysis. This dataset's rich and real-time nature allowed us to accurately calibrate and validate our proposed computational model. 
From September 2020 to January 2021, Germany witnessed significant events in its battle against the COVID-19 pandemic. In September, state authorities warned the public about rising infection rates and subsequently implemented partial lock-downs and increased restrictions on gatherings. Specifically, on October 28th, 2020, federal and state authorities agreed on a partial lockdown, limiting social contacts to two households. In December, facing a surge in the number of infectious cases, Germany implemented a stricter nationwide lockdown, shutting down non-essential businesses, schools, and cultural venues \cite{AP_News_2021}. During this period, the government initiated efforts to prepare for vaccination distribution, and on December 26, 2020, Germany began its vaccination campaign with the Pfizer-BioNTech vaccine. In January 2021, Germany expanded its vaccination efforts, targeting healthcare workers, elderly residents in care homes, and high-risk groups. While vaccination provided hope, new virus variants emerged in December 2020 and January 2021. $B.1.1.7$ variant was initially identified in the UK and is characterized by increased transmissibility, raising concerns about its potential impact on the pandemic's trajectory.
Meanwhile, the emergence of the $B.1.351$ variant in January 2021 sparked further worries about the efficacy of newly introduced vaccines. This new variant carries mutations that may impact the effectiveness of certain antibodies\cite{dw_news_2022}. Thus, this time window encompassed a complex interplay of lockdowns, vaccination challenges, and the evolving threat of virus variants, underscoring the rapid disease propagation and complex interplay of influential elements in its progression.

The hierarchical Bayesian modeling approach proposed in this paper predicts the number of infected individuals in Germany during the aforementioned timeline by incorporating critical events such as lockdowns, mutations, and vaccinations. Our results capture infection trends and estimate significant parameter variations in the average contact rate and probability of transmission variables. Furthermore, the model highlights the interplays of mutations and vaccinations and their subsequent effect on the transmission probability. Notably, the model demonstrates a strong correlation between the probability of transmission and infection rates while adjusting its predictions as new data regarding mutation, vaccination, and lockdowns becomes available.

Although we apply the model to Germany, our approach can be readily adapted to other countries or regions.

\section{Results}\label{sec:results_discussion}
\subsection{SARS-CoV-2 Progression in Germany in a Glance }\label{sec:covid_in_germany}
\begin{figure}[h!]
    \centering
\begin{tabular}{ccc}
    \subcaptionbox{Entire Period\label{et_i_pred}}{\includegraphics[width=0.5\textwidth]{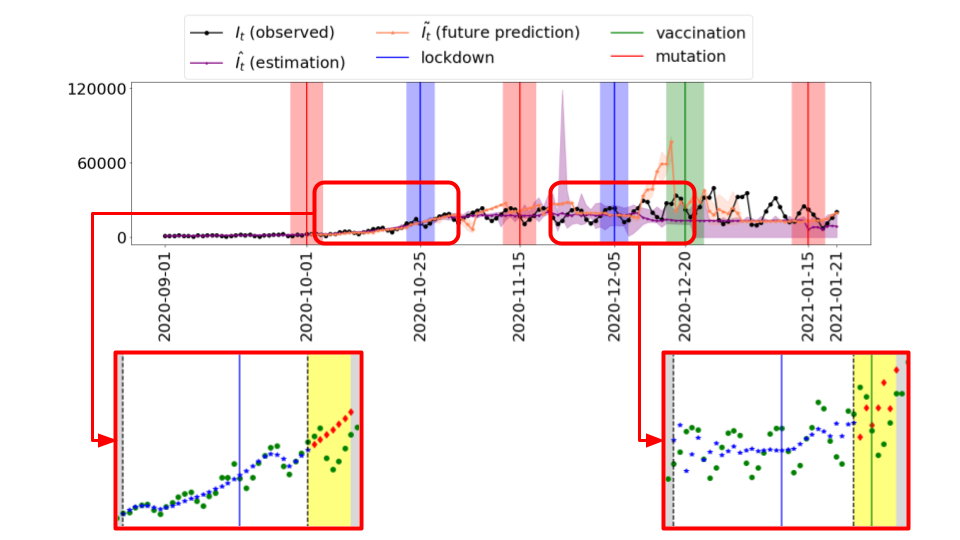}} &
    \subcaptionbox{Entire Period Prob\label{et_p_pred}}{\includegraphics[width=0.5\textwidth]{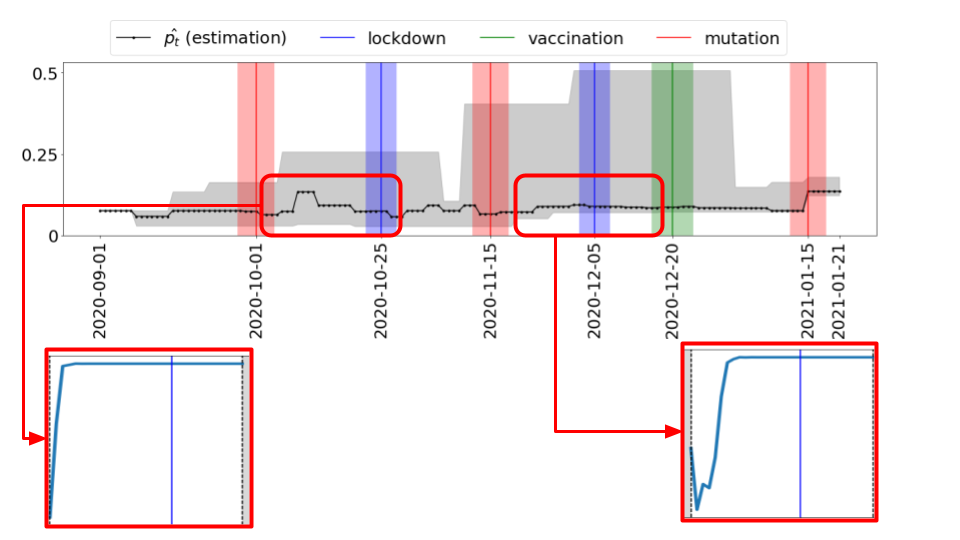}} 
\end{tabular}
\caption{Proposed model estimated values for $I_t$ and $p$ for 9/1/20 - 1/21/21 period.(\ref{et_i_pred}): The observed number of infected individuals is represented by black dots, while the median estimates and future forecasts are shown in purple and orange, respectively. The shaded regions indicate the $97.5^{th}$ percentile intervals for median estimates and future forecasts. The model accurately predicts infection trends, closely matching observed data until early December, with forecasts showing a rise in infections due to mutations before vaccination. A significant forecast increase between December $1^{st}$ and $20^{th}$ is corrected as the model incorporates newer data.
\ref{et_p_pred}: Illustrates the estimated probability transition rates over the same period. The black line shows the median probability transition, while the shaded area represents the $97.5^{th}$ percentile, indicating the uncertainty in these predictions. The probability transition rates exhibit increased uncertainty between the second virus mutation on November $15^{th}$ and vaccination on December $20^{th}$. This uncertainty is reflected in changing contributing factors, with the median estimate showing a fluctuating trend.}
\label{et_pred}
\end{figure}

To assess the progression of SARS-CoV-2 concerning the introduction of new variants and implementation of pharmaceutical and non-pharmaceutical in Germany, we focus on September $1^{st}$, 2020, to January $21^{st}$, 2021. This period explicitly exhibits the medium-term post-emergence feature in which hard-to-predict aspects of pathogen ecology and human behavior drive patterns. 
We performed the proposed hierarchical Bayesian online changing-point detection model coupled with the Poisson arrival process-based $SEIR$ model proposed in \cite{shahtori2022complex} (outlined in detail in \ref{sec:method}), which we previously published to address the dynamics of infectious disease spread. 
In the hierarchical Bayesian model, we used MCMC sampling to predict the number of infected individuals ($I_t$) and central parameters, including probability transition rate ($p_t$). \Cref{et_i_pred} and \ref{et_p_pred} illustrate the model's results. The critical periods, such as lockdowns, vaccinations, and mutations, are marked by blue, green, and red. The shaded areas around each event illustrate the transition intervals for the event to occur, in which they are sampled from a normal distribution during the transitions.

In \cref{et_i_pred}, the observed number of infected individuals is represented by black dots, while the median estimates and future forecasts are shown in purple and orange, respectively. The shaded regions indicate the $97.5^{th}$ percentile intervals for median estimates and future forecasts. These estimates are generated using the proposed framework, which processes the data in a sliding window fashion. Specifically, it first uses $31$ days of observed data (e.g., from September $1^{st}$ to October $1^{st}$) to model and predict the number of infected individuals (The two highlighted time periods provide a zoomed-in view of the predictions from the two sliding windows.). The data window is then shifted forward by 7 days (e.g., to September $8^{th}$ to October $8^{th}$), and the model is re-run using the newly observed data along with parameters estimated from the previous window. This process is repeated, and all estimates are collected. The median value of these estimates is then calculated to represent the number of infected individuals at each time point $t$. The median future forecast represents the central estimated number of infected individuals for seven days of future dates based on the parameters derived from a sliding window of 31 days of observed data. After estimating the model's parameters using the observed data, the model projects the future number of infections without relying on additional new observations. It generates multiple potential outcomes by sampling from the posterior distribution of the parameters and calculates the median of these predictions for each future time point. This median value provides a robust forecast of the expected number of infections, reflecting the most likely scenario while accounting for the uncertainties in the model parameters and disease dynamics.

As shown in \cref{et_i_pred}, the model effectively captures the trend in predicting the number of infections, with the median estimation closely aligning with the observed data until the beginning of December. From October onward, the future forecasts indicate a rise in infection numbers, corresponding to observed surges, particularly during the mutation and before vaccination. Notably, between December $1^{st}$ and December $20^{th}$, there is a significant jump in the future forecasts before the implementation of the vaccination. This jump is attributed to the unseen trend in the data. However, once the new sliding window incorporates more recent data, the model adjusts its predictions accordingly, demonstrating its ability to correct itself as new information becomes available. 

\cref{et_p_pred} illustrates the estimated probability transition rates over the same period. The black line shows the median probability transition, while the shaded area represents the $97.5^{th}$ percentile, indicating the uncertainty in these predictions. Notably, between the second virus mutation around November $15^{th}$ and the implementation of vaccination on December $20^{th}$, the contributing factors for the probability transition rate are consistently changing, resulting in a higher level of uncertainty. 
\begin{figure}[h!]
\centering
{\includegraphics[width=0.85\textwidth]{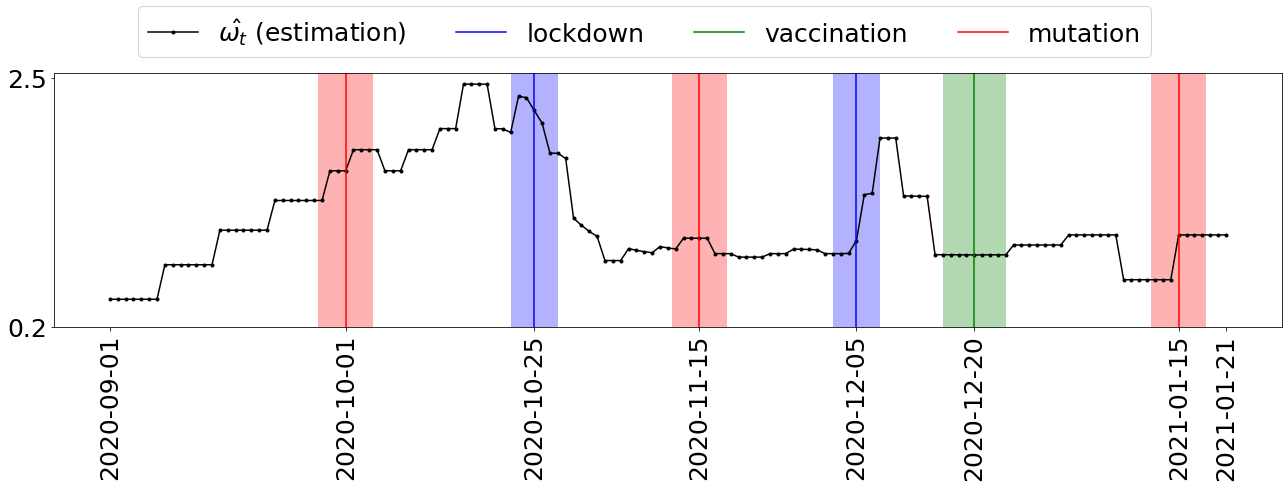}}
\caption{Proposed model estimated values for $\omega_t$ for 9/1/20 - 1/21/21 period. Before the October $25^{th}$, 2020 lockdown, the average contact rate rose following a mutation, but after the lockdown, it dropped and remained steady until a second lockdown around Christmas.}\label{fig:omega_t_prediction}
\end{figure}

\cref{fig:omega_t_prediction} illustrates the estimated average contact rate over the same period. As shown, before the implementation of lockdown around October $25^{th}$, 2020, the average contract rate increased, which coincided with the mutation introduction on October $1^{st}$, leading to an increase to $I_t$ (\cref{et_i_pred}). However, after the introduction of the first lockdown, the average contact drops and remains the same till around Christmas time in December, when the second lockdown gets introduced.

An important aspect of the results is the correlation and interplay between the number of infections ($I$), average contact rate ($\omega$), and the probability transition ($p$) over the given period. As shown in \cref{et_i_pred}, \cref{et_p_pred}, and \cref{fig:omega_t_prediction}, there is a notable correlation where increases in the probability transition precede and coincide with rises in the number of infections. For instance, after the mutation periods, there is a noticeable increase in the probability transition and/or average contact rate, followed by a rise in infection numbers. This relationship is consistent with the differential equations in \Cref{discrete_new_SEIR_eq} used in the model, where the infection rate is influenced by $p$, along with $\lambda_{EI}$ (the average number of exposed individuals who contracted the virus with probability p undergo a transition to state $E$). This correlation highlights the model's ability to capture the dynamics of disease transmission and the effects of critical events on infection trends. The strength of this modeling approach lies in its hierarchical Bayesian framework, which integrates prior knowledge and quantifies uncertainty. Additionally, the sliding window technique estimates model parameters and states in $31$-day increments, allowing the model to adapt to changing conditions and enhancing its short-term prediction accuracy.

\subsection{SARS-CoV-2 Progression Analysis Per Sliding Window and Error Estimation}\label{sec:covid_in_germany}

\begin{figure}[h!]
    \centering
\begin{tabular}{cccc}
    \subcaptionbox{9/1/20 - 10/1/20\label{fig:I_t_2020-09-01_2020-10-01}}{\includegraphics[width=0.45\linewidth]{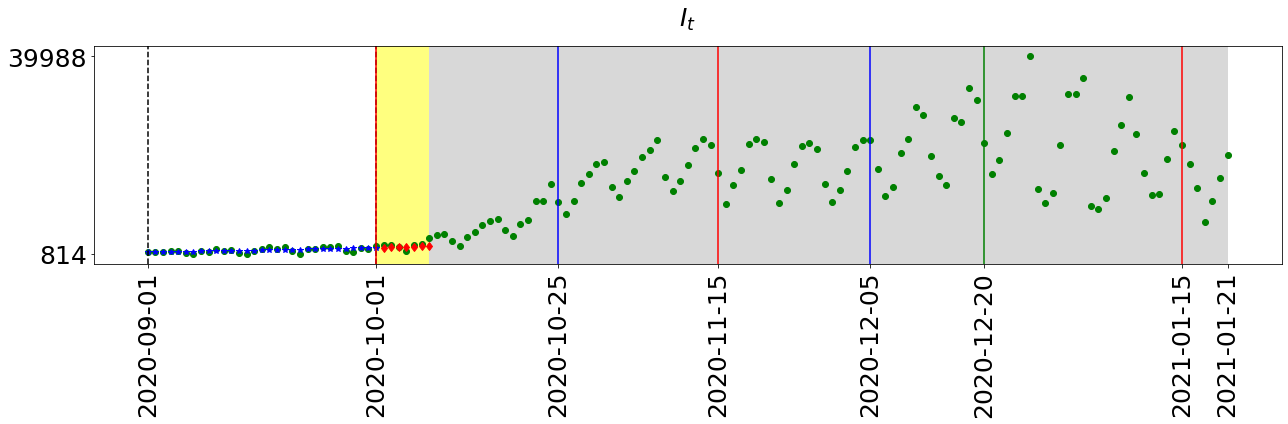}} 
    &
    \subcaptionbox{9/8/20 - 10/8/20\label{fig:I_t_2020-09-08_2020-10-08}}{\includegraphics[width=0.45\linewidth]{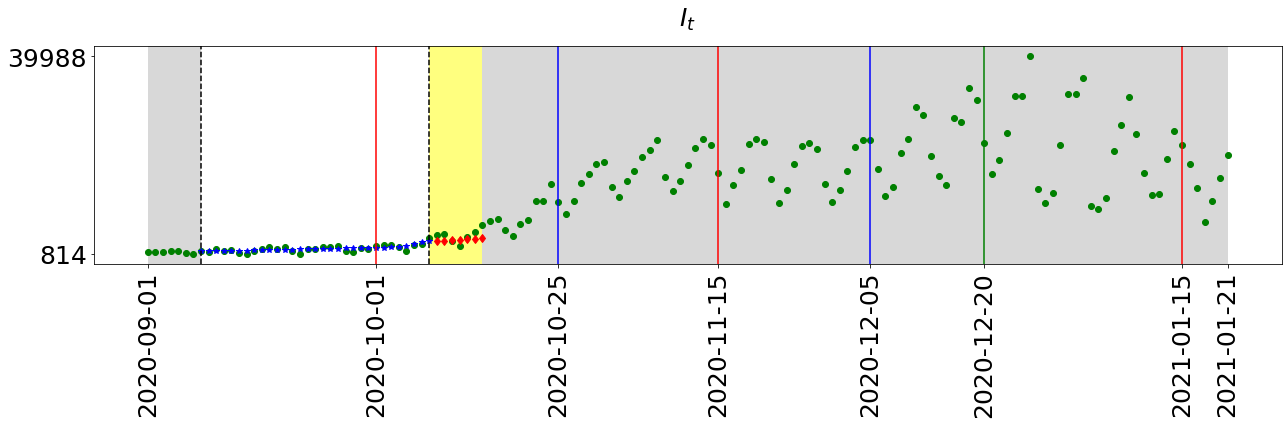}}
    \\
    \subcaptionbox{9/15/20 - 10/15/20\label{fig:I_t_2020-09-15_2020-10-15}}{\includegraphics[width=0.45\linewidth]{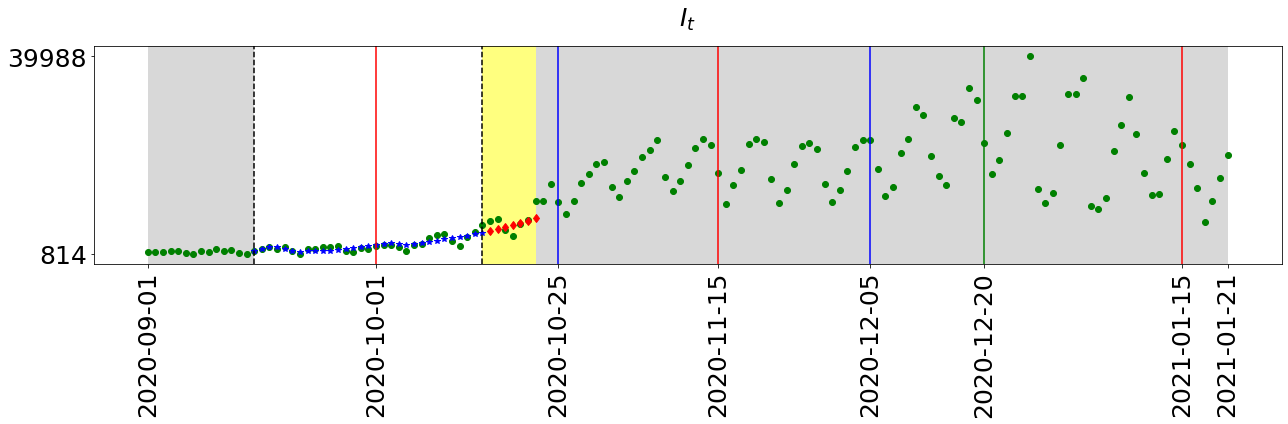}} 
    &
    \subcaptionbox{9/22/20 - 10/22/20\label{fig:I_t_2020-09-22_2020-10-22}}{\includegraphics[width=0.45\linewidth]{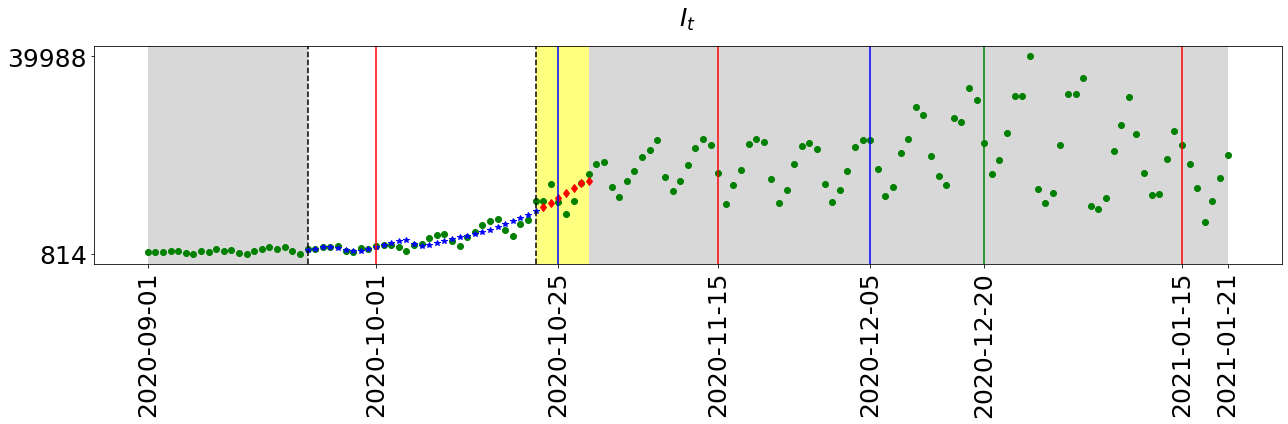}} 
    \\
    \subcaptionbox{9/29/20 - 10/29/20\label{fig:I_t_2020-09-29_2020-10-29}}{\includegraphics[width=0.45\linewidth]{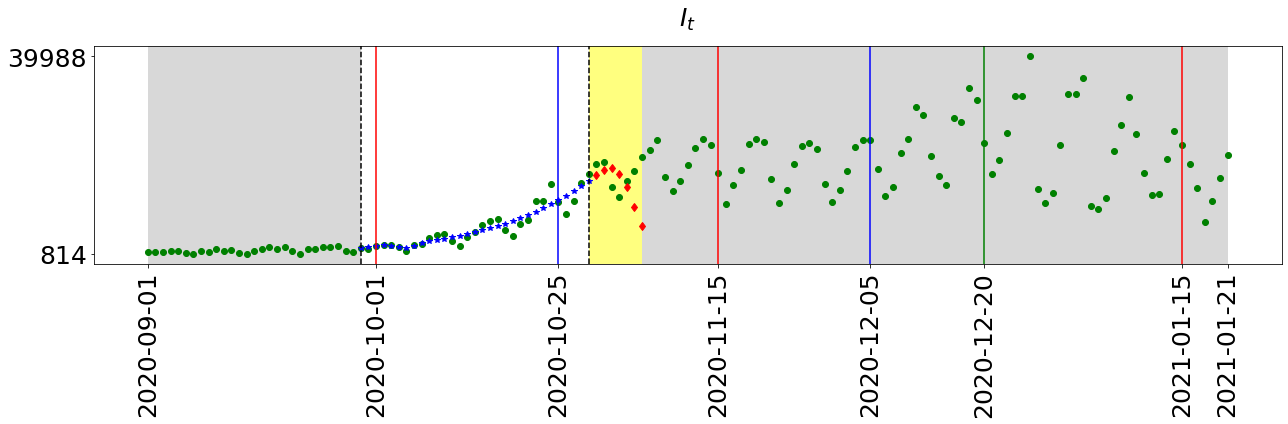}} 
    &
    \subcaptionbox{10/6/20 - 11/5/20\label{fig:I_t_2020-10-06_2020-11-05}}{\includegraphics[width=0.45\linewidth]{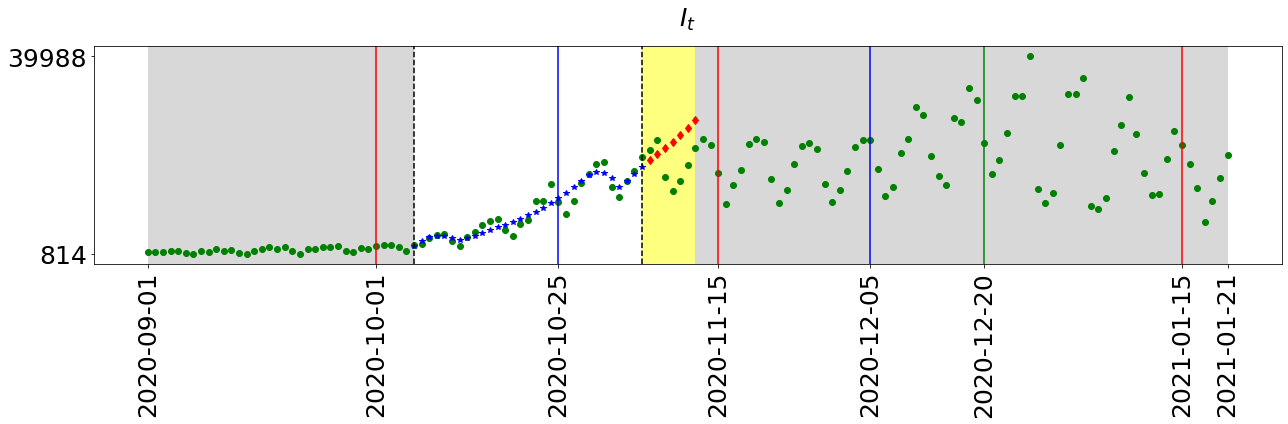}} 
    \\
    \subcaptionbox{10/6/20 - 11/5/20\label{fig:I_t_2020-10-13_2020-11-12}}{\includegraphics[width=0.45\linewidth]{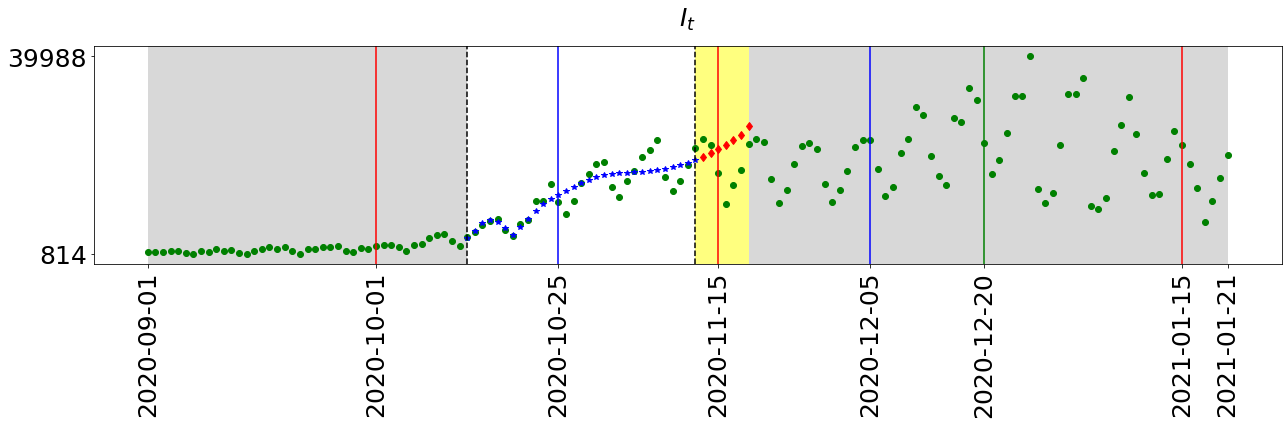}} 
    &
    \subcaptionbox{10/20/20 - 11/19/20\label{fig:I_t_2020-10-20_2020-11-19}}{\includegraphics[width=0.45\linewidth]{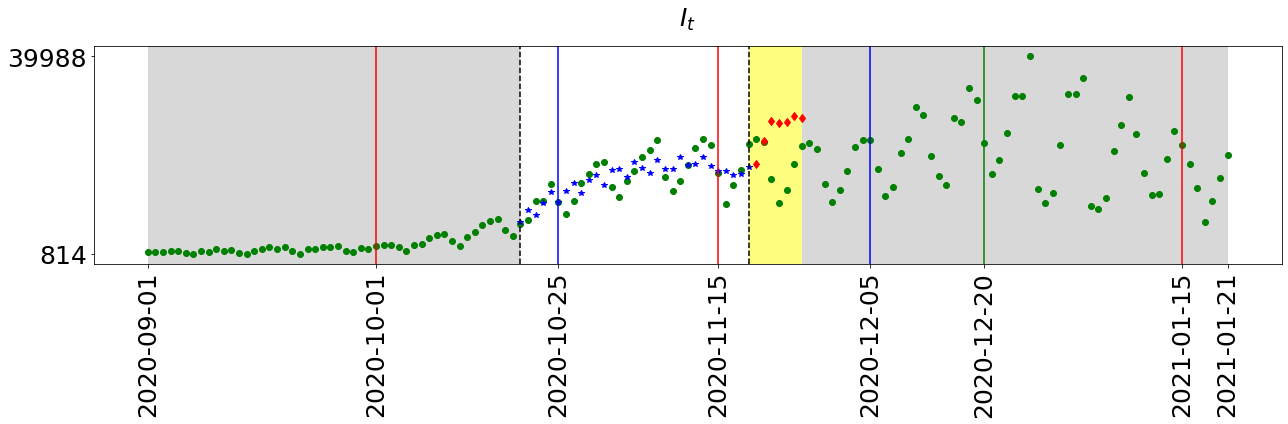}}
    \end{tabular}
\caption{\ref{fig:I_t_2020-09-01_2020-10-01}-\ref{fig:I_t_2020-10-20_2020-11-19} display the estimated values of $I_t$ using  Bayesian hierarchical changing point couple with 31 days sliding windows from 9/1/2020 - 11/19/2020 period. The green dots represent the observed number of infected individuals, while the blue markers show the model's estimates of infections. The white region indicates the 31-day window used to estimate model parameters, and the yellow area represents the week-long forecast based on these estimates. Vertical red lines mark the date of the mutation considered in the model, the green line shows when vaccination started, and the blue line represents the lockdown date. The sliding window approach iteratively adjusts predictions by incorporating new data every 7 days, generating median estimates to forecast future infection numbers. The results capture the overall trend in predicting the number of infections, with the median estimates closely matching the observed data and accurately reflecting fluctuations associated with key events during this period.}
\label{fig:sw_I_t_1}
\end{figure}

\begin{figure}[h!]
    \centering
\begin{tabular}{cccc}
    \subcaptionbox{10/27/20 - 11/26/20\label{fig:I_t_2020-10-27_2020-11-26}}{\includegraphics[width=0.45\linewidth]{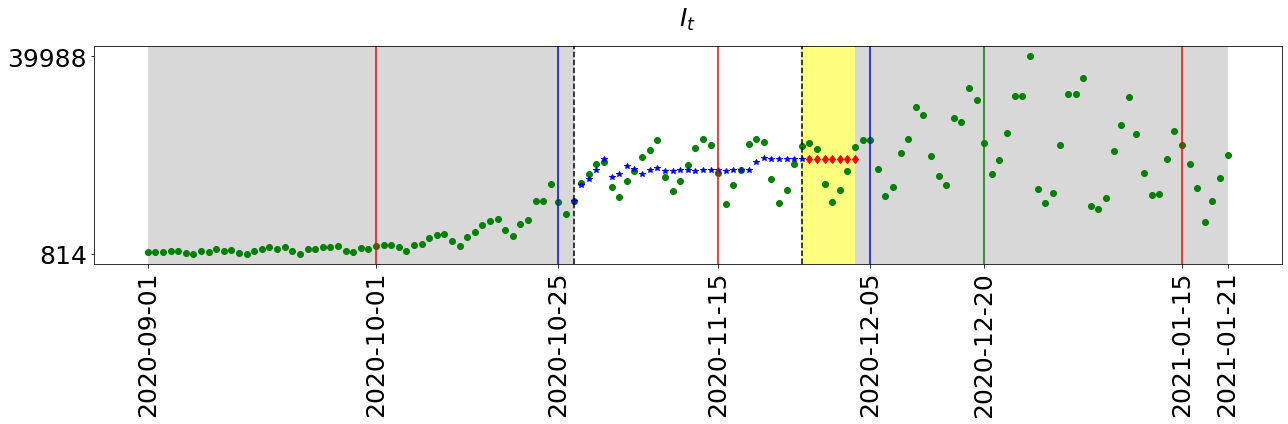}} 
    &
    \subcaptionbox{11/3/20 - 12/3/20\label{fig:I_t_2020-11-03_2020-12-03}}{\includegraphics[width=0.45\linewidth]{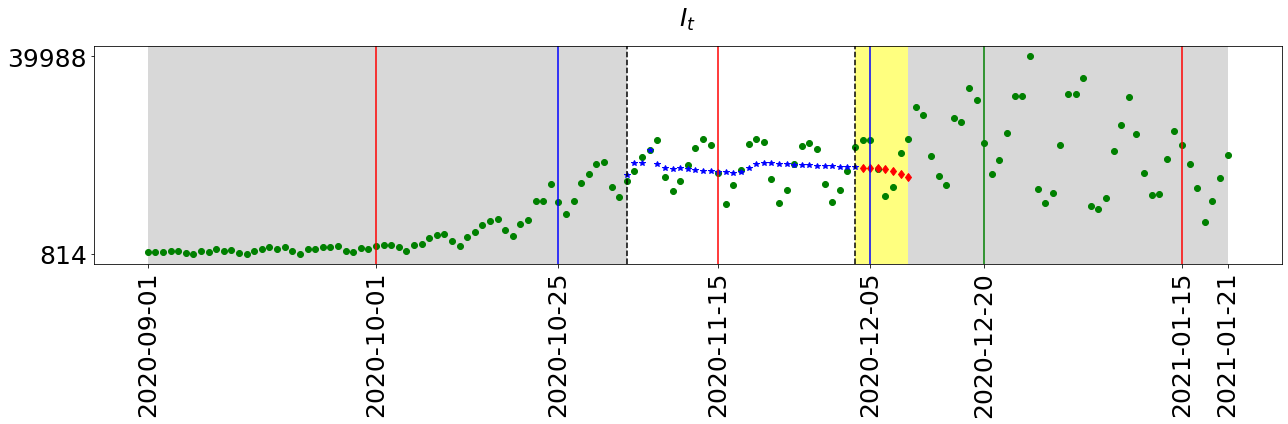}}
    \\
    \subcaptionbox{11/10/20 - 12/10/20\label{fig:I_t_2020-11-10_2020-12-10}}{\includegraphics[width=0.45\linewidth]{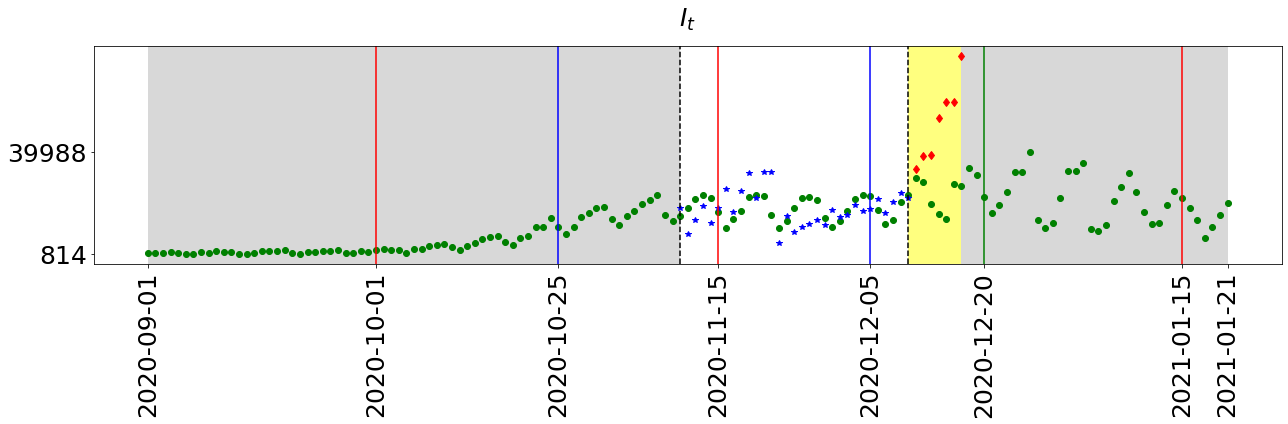}} 
    &
    \subcaptionbox{11/17/20 - 12/17/20\label{fig:I_t_2020-11-17_2020-12-17}}{\includegraphics[width=0.45\linewidth]{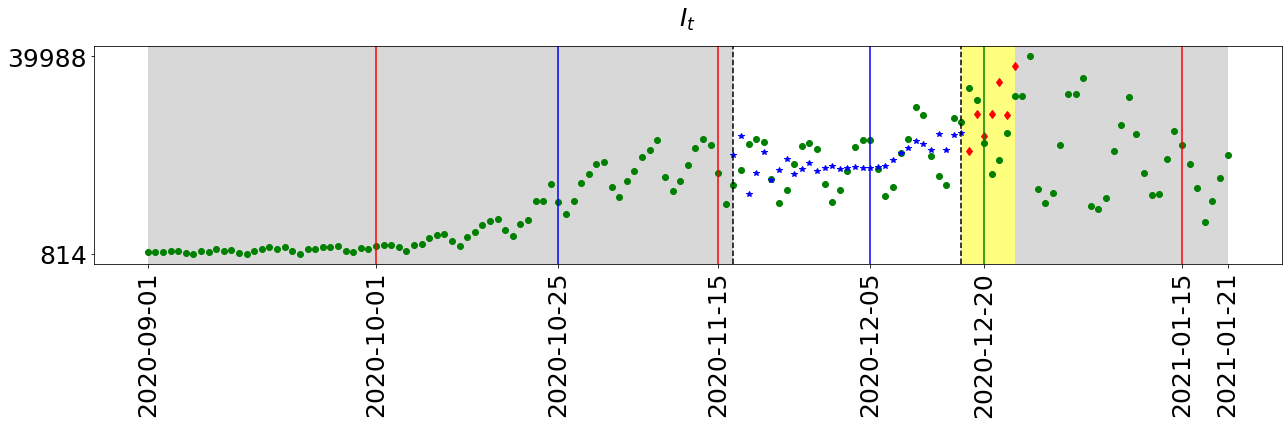}}
    \\
    \subcaptionbox{11/24/20 - 12/24/20\label{fig:I_t_2020-11-24_2020-12-24}}{\includegraphics[width=0.45\linewidth]{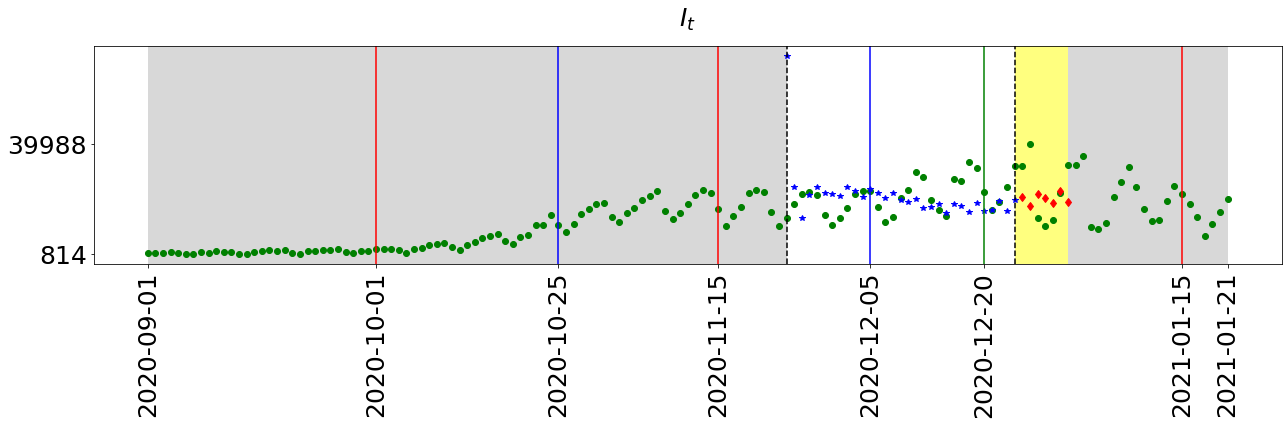}} 
    &
    \subcaptionbox{12/1/20 - 12/31/20\label{fig:I_t_2020-12-01_2020-12-31}}{\includegraphics[width=0.45\linewidth]{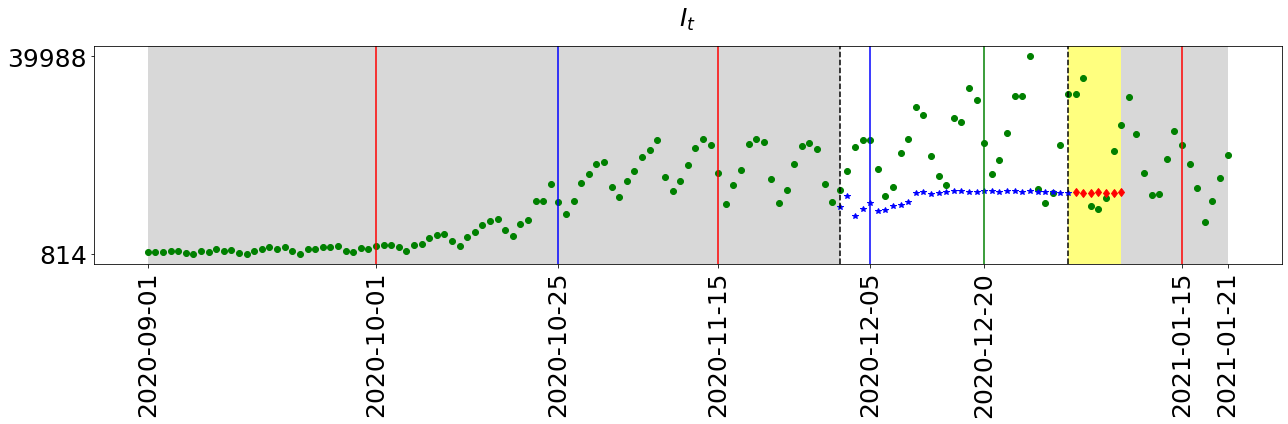}} 
    \\
    \subcaptionbox{12/8/20 - 1/7/21\label{fig:I_t_2020-12-08_2021-01-07}}{\includegraphics[width=0.45\linewidth]{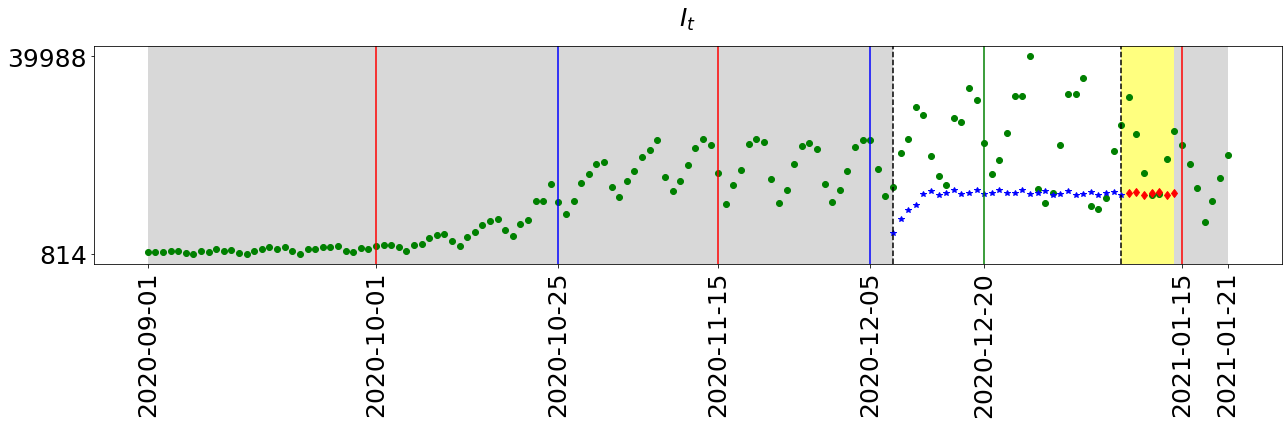}} 
    &
    \subcaptionbox{12/15/20 - 1/21/21\label{fig:I_t_2020-12-15_2021-01-14}}
    {\includegraphics[width=0.45\linewidth]{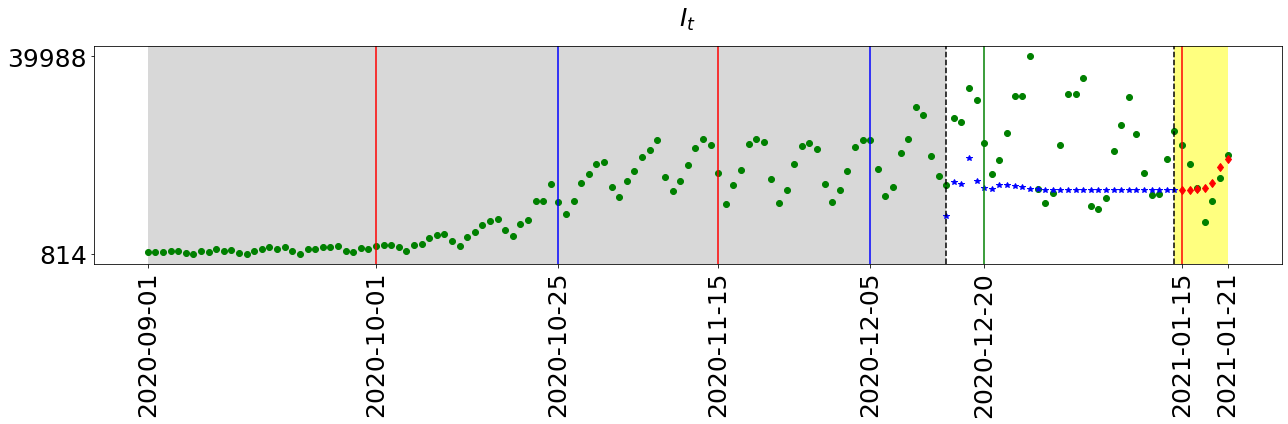}} 
    \end{tabular}
\caption{\ref{fig:I_t_2020-10-27_2020-11-26}-\ref{fig:I_t_2020-12-15_2021-01-14} display the estimated values of $I_t$ using  Bayesian hierarchical changing point couple with 31 days sliding windows from 10/27/2020 - 1/21/2021 period. The green dots represent the observed number of infected individuals, while the blue markers show the model's estimates of infections. The white region indicates the 31-day window used to estimate model parameters, and the yellow area represents the week-long forecast based on these estimates. Vertical red lines mark the date of the mutation considered in the model, the green line shows when vaccination started, and the blue line represents the lockdown date. The sliding window approach iteratively adjusts predictions by incorporating new data every 7 days, generating median estimates to forecast future infection numbers. The results capture the overall trend in predicting the number of infections, with the median estimates closely matching the observed data and accurately reflecting fluctuations associated with key events during this period.}
\label{fig:sw_I_t_2}
\end{figure}

The proposed algorithm provides a comprehensive framework for modeling and forecasting the spread of infectious diseases using a combination of differential equations, genetic algorithm-based prior selection, and a hierarchical Bayesian framework. This approach is designed to estimate model parameters and to forecast the number of infected individuals over a $31 days$ sliding window horizon, adapting dynamically to new data and changing epidemiological conditions. \cref{fig:sw_I_t_1} and \cref{fig:sw_I_t_2}, illustrates the predicted values for the number of infected individuals ($I_t$) across multiple 31-day sliding windows, with each window providing a 7-day forecast.  In both figures, the green dots represent the observed number of infected individuals, while the blue markers show the model's estimates of infections. The white region indicates the 31-day window used to estimate model parameters, and the yellow area represents the week-long forecast based on these estimates. Vertical red lines mark the mutation date considered in the model, the green line shows when vaccination started, and the blue line represents the lockdown date. The sliding window approach iteratively adjusts predictions by incorporating new data every 7 days, generating median estimates to forecast future infection numbers. The model effectively captures the underlying trends in infection rates, as the median estimations closely align with observed data, demonstrating its accuracy. Notably, the algorithm responds to critical epidemiological events, such as mutations and public health interventions. For example, it predicts a significant rise in infection rates following key mutation dates (October $1^{st}$, November $15^{th}$, and January $15^{th}$), which is indicative of the model's sensitivity to viral strain changes. The model also reflects the impact of public health measures; for instance, the contact rate ($\omega$) shows a marked decrease during lockdown periods (around October 25th and December 5th), corresponding with reduced social interactions due to imposed restrictions.

Moreover, the transmission probability ($p$) increases around mutation events, reflecting the enhanced transmissibility of new viral variants. The model accurately captures fluctuations in mutation rates ($m$) and vaccination efforts ($v$), particularly around significant milestones. For example, following the initiation of vaccinations on December 20th, the model predicts a subsequent reduction in transmission probability, highlighting the impact of vaccination on curbing the spread of the disease.

\begin{figure}[h!]
    \centering
    \subcaptionbox{\label{fig:error_estimation}}{\includegraphics[width=1\linewidth]{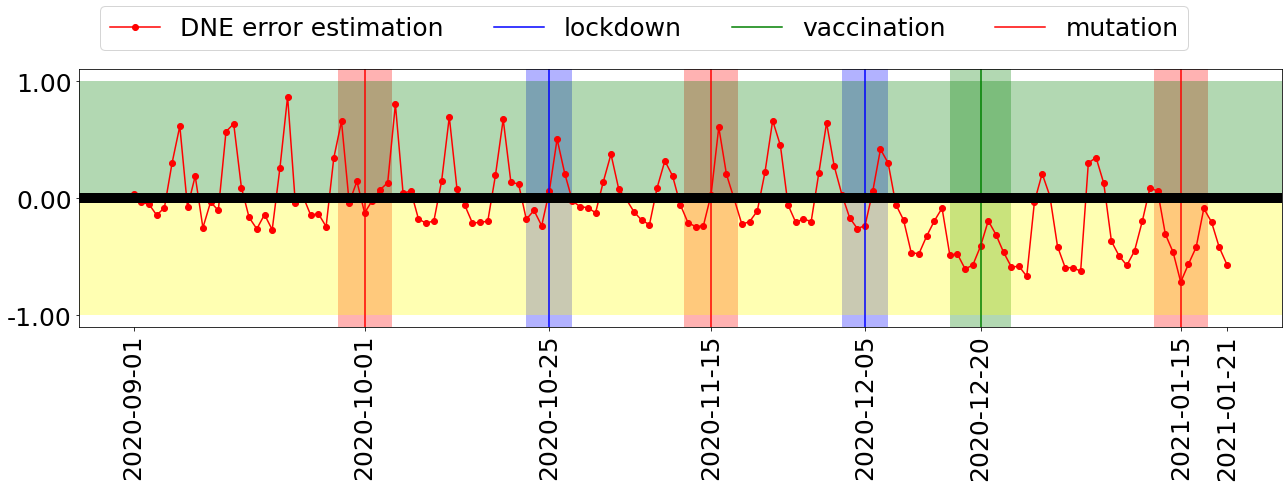}}
\caption{Directional error estimation ($DNE$) of $I_t$ for 9/1/20 -  1/21/2020. Around mutation dates, $DNE$ values increase, indicating that the model overestimates infection numbers due to challenges in adapting to new viral strains. During lockdown periods, $DNE$ decreases, often showing underestimation, while after the vaccination roll-out on December, $DNE$ stabilizes, suggesting better model alignment with observed data.}
\label{fig:error_estimation}
\end{figure}
We utilize directional error estimation ($DNE$) to monitor the model's errors effectively. This approach enables the identification of specific error patterns and their impacts on overall performance. \cref{fig:error_estimation} depicts the proposed model's DNE estimation, tracking its performance from September 2020 to January 2021. The $DNE$ is a metric for evaluating the accuracy of predicted infection individuals in the proposed hierarchical Bayesian framework by comparing the predicted number of infected individuals to the actual observed data. It calculates the difference between the predicted and observed values, then normalizes this error by dividing it by the observed value, thus expressing the error as a proportion of the actual value. A positive $DNE$ indicates that the model overestimated the number of infections (predicted higher than observed), while a negative $DNE$ reflects underestimation (predicted lower than observed). This provides a clear indication of the model’s performance across different phases of infection trends, helping assess whether the model systematically over or underestimates infection rates in certain periods, such as during a virus mutation or post-vaccination periods.
Around mutation dates (October $1^{st}$, November $15^{th}$, and January $15^{th}$), the DNE values exhibit clear spikes, indicating that the model overestimates infection numbers during these periods. This suggests that the model struggles to adjust promptly to the higher transmissibility associated with new viral strains. Conversely, during lockdown periods, such as on October 25th and December 5th, the DNE dips slightly below zero, pointing to an underestimation of infections. This trend likely reflects the reduced transmission resulting from social distancing and other containment measures. After the vaccination roll-out begins on December $20^{th}$, the DNE values become more stable, with fewer significant fluctuations. This stability suggests that the model aligns more closely with observed data as the vaccination campaign reduces the spread of infections, making the model's predictions more accurate.

\section{Discussion}\label{sec:discussion}
The proposed hierarchical Bayesian framework integrated with the Poisson arrival process-based SEIR model proposed in \cite{shahtori2022complex} effectively captures the progression of infectious disease dynamics, particularly in response to varying epidemiological factors such as mutation, immunity, and intervention measures. By leveraging a Poisson arrival process to model state transitions, the framework offers a nuanced representation of disease spread, where randomness in transition timings is accounted for. This method allows for more precise modeling of epidemic curves and enables the inclusion of real-world complexities such as varying contact rates and dynamic probabilities of infection. The results demonstrate that the model can accurately reflect the impact of interventions and changing virus characteristics, providing valuable insights into disease progression and control strategies.

A dynamic Bayesian inference framework enhances the model's robustness in dealing with noisy and incomplete data. Using a sliding window approach ensures that the model remains responsive to new information, such as emerging variants or shifts in public health measures. This adaptability is critical for real-time forecasting and decision-making, as it allows for continuous updating of parameter estimates and forecasts based on the most recent data. The hierarchical structure of the model also facilitates the integration of various sources of prior knowledge, which is particularly useful when data is scarce or highly uncertain. As demonstrated in the case study of Germany, this approach effectively captured the medium-term post-emergence dynamics of SARS-CoV-2, providing reliable estimates of infection rates and transmission probabilities during a period characterized by significant epidemiological changes.
However, the model's sensitivity to abrupt changes during critical periods necessitates continuous data updates and recalibration to maintain prediction accuracy. While the model effectively captured general trends and correlations, the exact timing and magnitude of infection peaks varied, highlighting the inherent uncertainty in forecasting and the need for ongoing refinement. Overall, this modeling approach provides a comprehensive framework for understanding and predicting infection trends, contributing to more effective disease control and prevention efforts.

\section{Methods}\label{sec:method}
\subsection{Data collection and processing}\label{sec:data_collection}
We used data provided by the Johns Hopkins University Center for Systems Science and Engineering (JHU CSSE) dashboard \cite{dong2020interactive}. The data becomes available through a collaborative effort involving global health organizations, governments, and institutions that collect, verify, and report COVID-19 cases. We use COVID-19 data from Germany, spanning September $1^{st}$, 2020, to January $21^{st}$, 2021. We utilized the extensive dataset provided by Johns Hopkins University’s COVID-
19 repository as a cornerstone for our analysis. The rich
and real-time nature of this dataset allowed us to calibrate and validate our computational model accurately. We specifically target the timeframe as a pivotal juncture within the COVID-19
pandemic timeline in Germany, which represents the characteristic of medium post-emergence and contains 143 distinctive data points. For days reported as zero, we replaced those data points with the last reported number of cases to avoid model hallucination.   

\subsection{Integrated Framework for Dynamic Bayesian Inference of Infectious Disease Spread}\label{sec:bsw_model}
\begin{figure}[h!]
    \centering
    \includegraphics[width=1\linewidth]{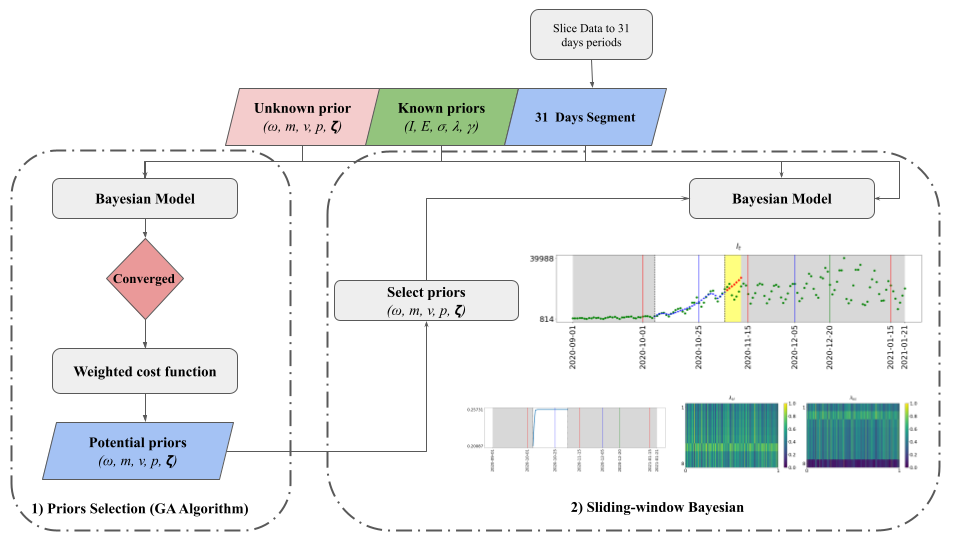}
    \caption{Proposed Framework for Bayesian Inference of Infectious Disease Dynamics with Genetic Algorithm-based Prior Selection and Sliding-Window Forecasting: It illustrates the workflow of the proposed algorithm for estimating and forecasting the spread of infectious disease using a Bayesian model informed by differential equations and a genetic algorithm (GA) for prior selection.}
    \label{fig:bsw_algo}
\end{figure}

The proposed algorithm provides a comprehensive framework for modeling and forecasting the spread of infectious diseases using a combination of differential equations, genetic algorithm-based prior selection, and a hierarchical Bayesian framework. This approach is specifically designed to estimate model parameters robustly and to forecast the number of infected individuals over a medium-term horizon, adapting dynamically to new data and changing epidemiological conditions. As illustrated in \cref{fig:bsw_algo}, the workflow integrates several interdependent components to achieve accurate and reliable disease modeling and forecasting.

The algorithm begins with a Poisson-based $SEIR$ (Susceptible-Exposed-Infectious-Recovered) model, Eq.\eqref{discrete_new_SEIR_eq}, proposed by \cite{shahtori2022complex}, which describes the transitions between different disease states using a set of differential equations. These equations incorporate various factors influencing the disease dynamics, such as transmission rate, mutation rate, vaccination rate, and coefficients affecting the probability of progression from the exposed to the infectious state. The rate at which susceptible individuals become exposed is governed by the average contact rate $\omega$, and once exposed, individuals may progress to the infectious state with a probability $p$ and transition rate $\lambda_{EI}$ or revert to the susceptible state at rate $\lambda_{ES}$. Infectious individuals then transition to the recovered state at rate $\lambda_{IR}$ after an infectious period $\gamma$. The probability $p$ of transitioning from exposed to infectious is influenced by time-varying factors such as virus mutation $m(t)$, waning immunity, and vaccination efforts $v(t)$. These dynamics are captured through differential equations, enabling the model to account for stochastic variations in disease spread. More details about the implemented mechanistic model can be found in the section. \ref{sec:pseir_model}. 

Given the complexity of these dynamics, the algorithm uses a genetic algorithm (GA) to identify suitable initial priors for parameters that lack reliable information, such as $\omega$, $v$, $m$, and coefficients $\zeta_0$, $\zeta_1$, and $\zeta_2$ that governs dynamics of $p$. The GA optimizes a weighted cost function by iteratively searching the parameter space and selecting and recombining parameters to minimize discrepancies between model predictions and observed data. This iterative process continues until convergence is achieved, resulting in a set of potential priors for the Bayesian model. These priors serve as the initial estimates for the Bayesian framework, which is crucial for ensuring the model's convergence and accuracy. More details about the selected priors can be found in the section. \ref{sec:gen_algo}. 

Once the priors are established, the hierarchical Bayesian framework is implemented to refine these parameter estimates and forecast disease dynamics. The model uses Markov Chain Monte Carlo ($MCMC$) sampling to generate samples from the posterior distribution of parameters and state variables. This approach integrates prior knowledge, observed data, and model structure to provide robust estimates of the uncertainties associated with the number of infected individuals ($I_t$) and model parameters. By accounting for noisy and incomplete data, the Bayesian framework enhances the model's ability to produce accurate forecasts and to update these forecasts as new information becomes available. The Bayesian model adapts dynamically to changes in pathogen dynamics, such as mutations and immunity development, and to shifts in human behavior, such as changes in social distancing measures.

A sliding-window technique is implemented to ensure the model continuously adapts to new incoming data. Rather than analyzing the entire dataset simultaneously, the data is divided into overlapping segments, each spanning $31$ days. For each segment, the Bayesian model is initialized with one of the priors obtained by the GA and updated with observed data. It then forecasts the number of infected cases for the next $7$ days. After each forecast, the window shifts forward by $7$ days, and the process is repeated with the new data segment. This iterative procedure allows the model to adapt to changes in the data, providing more accurate forecasts even in the presence of rapidly evolving epidemiological conditions. This method enables continuous updating of parameter estimates and forecasts, making the model highly adaptable to real-time changes in disease dynamics.

\subsection{Mechanistic model}\label{sec:pseir_model}
We consider a time-discrete version of the Poisson arrival process-based $SEIR$ model proposed in \cite{shahtori2022complex}. The authors previously published this model to delineate transitions between various states, allowing the tracking of the exposed population effectively. The proposed model, based on random event-based Poisson arrivals, offers a comprehensive framework for understanding disease spread when the exposed state is intermediate between susceptibility and infectiousness and delays in implementing mitigation strategies are inevitable. In this model, the transitions between $E \rightarrow I$ and $I \rightarrow R$ are considered an arrival Poisson point process, meaning an individual in the host population arrives at a new state (given the health status) within a predefined period. We can model this behavior using the Poisson point process concept because the average transition period is known, but the exact arrival time to a new state is random \cite{shahtori2022complex}.

In this model, a susceptible individual who encounters an infectious person leaves the susceptible group ($S$) with a rate $\omega$. The newly exposed individuals (\textit{i.e.} $E_{new} = \omega S^t I^t$) at time $t$, are considered non-symptomatic and non-infectious. Notably, only a fraction of exposed individuals become infectious. The exposed individuals contracting the disease with probability $p$ move to the infectious state $I$ with rate $\lambda_{EI}$ after incubating the disease for $\sigma$ unit time (\textit{i.e.} $I_{new} = p \lambda_{EI} E^{t-\sigma}$). The exposed individuals who do not contract the disease after $\sigma$ unit time return to the susceptible population with rate $\lambda_{ES}$. Then, infected individuals enter the recovered/dead state ($R$) with rate $\lambda_{IR}$ after $\gamma$ unit time.  

This framework formulates the dynamic changes of various
sub-clusters of the host population based on their exposure
to the virus when the exposed group is intermediate between susceptibility and infectiousness. Furthermore, the proposed model allows for constructing $p$ as a function of contributing factors. Specifically, the proposed model incorporates $p(t)$ as part of the differential equation by Integrating contributing factors: virus mutation $m(t)$, waning immunity, and the population's immunity resilience against the virus boosted by vaccination efforts $v(t)$. \Cref{new_SEIR_eq} represents the computational framework for the proposed model.
\begin{equation}
\begin{split}
\frac{dS}{dt} & = -\omega \frac{S^{t}I^{t}}{N}
                   + \lambda_{ES} E^{t - \sigma}, 
                   \\
\frac{dE}{dt} & = \omega \frac{S^{t}I^{t}}{N}
                  - \lambda_{ES} E^{t - \sigma} 
                  - p \lambda_{EI} E^{t - \sigma}, 
                  \\
\frac{dI}{dt} & = p \lambda_{EI} E^{t - \sigma} 
                  - \lambda_{IR} I^{t - \gamma}, 
                  \\
\frac{dR}{dt} & = \lambda_{IR} I^{t - \gamma}, \\
\frac{dp}{dt} & = -[\delta + \zeta_0 v^{t}] \zeta_1 p^t 
                  + \zeta_2 m^{t} 
\end{split}
\label{new_SEIR_eq}
\end{equation}

In \Cref{new_SEIR_eq}, $\zeta_1$ represents the overall rate at which natural immunity is produced against the virus in the host population. $\zeta_2$ is the rate at which the virus spreads given the circulating variant at time $t$, and $\zeta_3$ represents how well the vaccination efforts are implemented. Further background about the reformulation of the SEIR model can be found in \cite{shahtori2022complex}.  

Since our dataset is discrete in time $\Delta t = 1$, we discretize the model using the forward Euler method assuming $\Delta t = 1$ such that,
\begin{equation}
\begin{split}
S^{k + 1} &= S^{k}  
                   - \omega \frac{S^{k}I^{k}}{{N}} 
                   + \sum_{\sigma=1}^{7} \lambda_{es}^{k - \sigma} E^{k - \sigma }, \\
E^{k + 1} &= E^{k} 
                  - \omega \frac{S^{k}I^{k}}{{N}} 
                  - \sum_{\sigma=1}^{7} \lambda_{es}^{k - \sigma} E^{k - \sigma } 
                  - \sum_{\sigma=1}^{7} p \lambda_{ei}^{k-\sigma} E^{k - \sigma },  \\
I^{k + 1} &= I^{k} 
                  + \sum_{\sigma=1}^{7} p \lambda_{ei}^{k-\sigma} E^{k - \sigma }
                  - \sum_{\sigma=1}^{11} p \lambda_{ir}^{k-\gamma} I^{k - \gamma },  \\
R^{k + 1} &= R^{k} 
                 + \sum_{\sigma=1}^{11} p \lambda_{ir}^{k-\gamma} I^{k - \gamma },
                 \\
p^{k+1} &= [\delta -\zeta_0 v^{k}] \zeta_1 p^{k} + \zeta_2 m^{k}, 
\end{split}
\label{discrete_new_SEIR_eq}
\end{equation}
$I^k$ represents the number of currently active infected people, while $I^{k + 1}$ is the number of new infections that will eventually be reported. In this model, we simplified the dynamics by excluding the return of recovered individuals to the susceptible state. This simplification was made to streamline the model and focus on the primary interactions between susceptible, exposed, infectious, and recovered individuals. 

\subsection{Bayesian Framework and Initial condition}\label{sec:gen_algo}
We construct the Bayesian hierarchical framework as follows.
\begin{equation}
p(\Omega, \Theta, \chi | I) \propto p(I | \Omega, \Theta, \chi) p(\Omega, \Theta, \chi)
\label{hierarchical_bayesian_eq}
\end{equation}
In \Cref{hierarchical_bayesian_eq}, $P(\Omega, \Theta, \chi | I)$ denotes the posterior distribution, derived as the product of the likelihood $P(I | \Omega, \Theta, \chi)$ and the prior $P(\Omega, \Theta, \chi)$. Additionally, $\Omega$ represents the observed and unobserved states of the system, $I$ and $\{S, E, R\}$ respectively at each step. $\Theta$ encapsulates the model parameters, which includes $\{\omega, p, \gamma, \sigma, \lambda_{es}, \lambda_{ei}, \lambda_{ir}\}$. Lastly, $\chi$ represents the hyper-parameters governing the prior distributions. The likelihood of observing the infected cases \( I \) is modeled using a Poisson distribution as follows
\begin{equation}
p(I^k | I^{1:k-1}, \Theta, \chi) \propto \text{P}(I^k | \lambda[f(\Omega^{k-1}, \Theta)])
\end{equation}
where $\lambda[f(\Omega^{k-1}, \Theta)]$ denotes the model-predicted mean for the number of infected individuals ($I$) at time step $k$, calculated based on the state transition function $f(\Omega^{k-1}, \Theta)$ defined in \Cref{discrete_new_SEIR_eq} and the updated parameters $\Theta$. Likelihood quantifies how well the model explains or fits the observed data. In \Cref{discrete_new_SEIR_eq}, the transitions between $S \rightarrow E \rightarrow I$ and $I \rightarrow R$ states are defined as an arrival Poisson point process, meaning an individual in the host population arrives at a new state (given the health status) within a predefined period. This behavior is modeled using the Poisson point process concept because the average transition period is known, but the arrival time to a new state is random. In the statistical mechanistic model, the transition rates are defined as the average number of arrivals (events) given the transition period. As a result, we define the likelihood function as Poisson distribution to account for the uncertainty of parameters and quantify the similarity between the model outcome and partially observed data (state $I^{obs}$). 

The choice of the prior distribution is a crucial step in the Bayesian framework as it encapsulates the prior beliefs about the parameters before observing the data. We used normal distribution during the medium-term post-emergence for variables where prior information about the parameter’s central tendency is available. This includes variables such as the incubation period ($\sigma$) and recovery period ($\gamma$) as defined in \cref{prior_normal_dist}. 
\begin{equation}
\begin{split}
\kappa_{\mathcal{N}} &\propto \mathcal{N} (\mu_{\kappa_{\mathcal{N}}},\sigma_{\kappa_{\mathcal{N}}}),\\
\kappa_{\mathcal{N}} &\in \{\sigma, \gamma\}\\
\end{split}
\label{prior_normal_dist}
\end{equation}
For other variables that are positively skewed and can only take positive values such as $\zeta_0$, $\zeta_1$, $\zeta_2$, $\lambda_{ei}$ and $\lambda_{ir}$ we used Log-normal distribution and Gamma distribution for $\lambda_{ei}$ which is a strictly positive quantity as defined in \cref{prior_lognormal_dist}. 
\begin{equation}
\begin{split}
\kappa_{\textit{L}}  &\propto \textit{LogN} (\mu_{\kappa_L}, \sigma_{\kappa_{\textit{L}}}),\\
\kappa_{\textit{L}} &\in \{\zeta_0, \zeta_1, \zeta_2, \lambda_{ei}, \lambda_{ir}\}\\
\kappa_{\textit{G}}  &\propto \mathcal{G} (\mu_{\kappa_{\textit{G}}}, \sigma_{\kappa_{\textit{G}}}),\\
\kappa_{\textit{G}} &\in \{\lambda_{es}\}\\
\end{split}
\label{prior_lognormal_dist}
\end{equation}
To accommodate outliers, which might occur in epidemiological models due to unexpected events or fluctuations in $p$, $I$ and $E$, we used Half student-T distribution to provide robustness against outliers as defined in \cref{prior_halft_dist}. 
\begin{equation}
\begin{split}
\kappa_{t}  &\propto t_{\textit{half}} (v_{\kappa_{t}}, \sigma_{\kappa_{t} }),\\
\kappa_{\textit{t}} &\in \{E_0, I_0, p_0\}\\
\end{split}
\label{prior_halft_dist}
\end{equation}

Lastly, we used log normal distribution with change-points for average contact rate ($\omega$), mutation ($m$) and vaccination ($v$) and for the timing of change points, we chose
normally distributed priors, as defined in \cref{prior_changepoints_dist}. Change-points proposed by \textit{Dehning et. al} \cite{dehning2020inferring}, models a time-dependent parameter using a framework that incorporates change points. This approach allows the parameter to exhibit shifts at specific times, with the transitions between different states being represented by sigmoidal functions. The sigmoid function is a mathematical tool that produces an S-shaped curve, which is used to smoothly interpolate between different values at each change point. The function starts with an initial value for the defined variable and then incrementally adjusts it based on the change points. Each change point is characterized by a sigmoid function that models the transition from one state to another, ensuring that changes occur in a gradual, rather than abrupt, manner. Specifically, the sigmoid function ensures that the defined variable changes smoothly over time at each change point. The steepness of the transition is controlled by the scaling factor in the sigmoid function, which adjusts how quickly the parameter shifts from one value to the next. By summing all these sigmoid-modified transitions, the function creates a composite time-dependent parameter that captures the cumulative effect of all change points. This framework is particularly useful for modeling scenarios where parameters evolve due to significant, but gradual, changes over time, such as in epidemiological studies where the spread of a disease might change due to interventions or other factors.

\begin{equation}
\begin{split}
\kappa_{\textit{C}}  &\propto \textit{LogN} (\mu_{\kappa}, \sigma_{\kappa}),\\
\kappa_{\textit{C}} &\in \{\omega, m, v\}\\
\end{split}
\label{prior_changepoints_dist}
\end{equation}

To initialize the Bayesian model for the first segment of data, that is Sep $1^{st}$ - Oct $1^{st}$, and select the central points for model variables such as $\omega$, $v$, $m$, and $\zeta_0$, $\zeta_1$, $\zeta_2$, for which no reliable information is available for model initialization, genetic algorithm (GA) is utilized. This approach aims to explore the space of possible priors for these parameters. Specifically, the GA searches for suitable values within defined bounds for each parameter. The bounds are set to ensure comprehensive exploration of feasible values, allowing the GA to efficiently search through this space and propose a diverse set of potential priors. For instance, the space for $\omega$ change-points, i.e. lock-downs, is defined as $[0.1, 0.9]$ and the initial condition $\omega_0$ is within $[0.2, 0.6]$. For the two mutation change-points, as well as the initial condition $m_0$, the bound is set to $ (0, 0.8]$. For the vaccination change-point, i.e. implementation of first batch of vaccination, the bound is $0, 0.2]$, and for the initial condition $v_0$, the space is $[0, 0.00001]$ to minimize the impact of vaccination in estimation. Lastly, the bounds for the initial conditions $p$, $\zeta_0$, $\zeta_1$, and $\zeta_2$ are defined as $(0, 1]$.

The GA operates by evaluating different sets of initial priors through a fitness function that assesses the quality of the predictions made by the Bayesian model. This function calculates a weighted cost based on how well the predicted new cases align with observed data, incorporating both time decay and value scaling of the data points. Through iterative optimization, the GA refines the priors to minimize this weighted cost, ultimately identifying promising candidates for initializing the Bayesian framework. The advantage of using this framework lies in its ability to leverage the GA's search capabilities to find a range of plausible priors rather than relying on a single, potentially arbitrary choice. This method provides a selection of priors that can be used to initialize Bayesian inference, accommodating the inherent uncertainty in prior selection. By exploring a broad parameter space, the framework enhances the robustness and flexibility of the Bayesian modeling process, potentially leading to more accurate and reliable inference outcomes. Table. \ref{tab:prior_distributions} summarize the selection of priors and initial condition for mechanistic \cref{discrete_new_SEIR_eq}. 
\begin{table}[h!]
\centering
\caption{Prior Distributions for Model Variables}
\begin{tabular}{@{}lll@{}}
\toprule
\textbf{Variable}       & \textbf{Description}                                & \textbf{Distribution and Parameters} \\ \midrule
\( \omega_0 \)          & Initial condition for contact rate  & \( \textit{LogN}(3.44 \times 10^{-1}, 0.05) \) \\
\( \omega_1 \)          & First lockdown change point event      & \( \textit{LogN}(4.06 \times 10^{-1}, 0.05) \) \\
\( \omega_2 \)          & Second lockdown change point event      & \( \textit{LogN}(3.76 \times 10^{-1}, 0.05) \) \\
\( v_0 \)               & Initial condition for vaccination rate    & \( \textit{LogN}(5.98 \times 10^{-8}, 0.5) \) \\
\( v_1 \)               & Vaccination change point event                          & \( \textit{LogN}(1.44 \times 10^{-2}, 0.05) \) \\
\( m_0 \)               & Initial condition for mutation rate       & \( \textit{LogN}(3.02 \times 10^{-1}, 0.5) \) \\
\( m_1 \)               & First mutation change point event                        & \( \text{LogN}(3.63 \times 10^{-1}, 0.05) \) \\
\( m_2 \)               & Second mutation change point event                     & \( \text{LogN}(3.99 \times 10^{-1}, 0.05) \) \\
\( m_3 \)               & Third mutation change point  event                      & \( \text{LogN}(4.21 \times 10^{-1}, 0.05) \) \\
\( p_0 \)               & Initial condition for probability tranmission           & \( t_{\text{half}}(4.99 \times 10^{-3}, 4) \) \\
\( I_0 \)               & Initial number of infected individuals             & \( t_{\text{half}}(1200 , 4) \) \\
\( E_0 \)               & Initial number of exposed individuals              & \( t_{\text{half}}(100 , 4) \) \\
\( \delta \)            & Probability transmission scaling factor                & \(0.4 \) \\
\( \zeta_0 \)           & Initial condition for \(\zeta_0\)                     & \( \text{LogN}(9.65 \times 10^{-1}, 0.5) \) \\
\( \zeta_1 \)           & Initial condition for \(\zeta_1\)                     & \( \text{LogN}(6.89 \times 10^{-1}, 0.5) \) \\
\( \zeta_2 \)           & Initial condition for \(\zeta_2\)                    & \( \text{LogN}(3.13 \times 10^{-1}, 0.5) \) \\ 
\( \lambda_{es} \)      & Rate parameter for \( \lambda_{es} \)               & \( \text{LogN}(\text{log}(8), 0.5) \) \\ 
\( \lambda_{ei} \)      & Rate parameter for \( \lambda_{ei} \)               & \( \mathcal{G}(\text{log}(8), 0.5) \) \\ 
\( \lambda_{ir} \)      & Rate parameter for \( \lambda_{ir} \)               & \( \text{LogN}(\text{log}(22), 0.1) \) \\ 
\bottomrule
\label{tab:prior_distributions}
\end{tabular}
\end{table}

\subsection{Model Initialization \& MCMC Sampling}\label{sec:model_init}
We used the $advi+adapt\_diag$ initialization method implemented in $PyMC4$ to approximate the posterior distribution of model parameters based on the observed data for all sliding windows. The $advi+adapt\_diag$ method approximates the posterior distribution of model parameters, which combines two techniques: automatic differentiation variational inference ($ADVI$) and the adaptation scheme $adapt\_diag$ . The $ADVI$  technique aims to discover a simpler distribution, typically belonging to the Gaussian family, that closely approximates the actual posterior distribution while minimizing the relative entropy. A diagonal mass matrix is part of the $ADVI$ optimization process to balance the parameter space exploration. $adapt\_diag$  method creates a dynamic method to adapt the diagonal elements of the mass matrix during the optimization process. The adaptation of the the $adapt\_diag$  method helps in scaling and controlling the parameter space exploration.

For each sliding window, we performed Bayesian estimation using $500$ tuning steps to adjust sampling parameters for better convergence, followed by drawing $500$ posterior samples. Specifically, each chain underwent 500 tuning steps using NUTS, although these were not recorded. NUTS is a gradient-based Hamiltonian Monte Carlo (HMC) sampling algorithm that simulates the behavior of particles in a physical system to explore the posterior distribution efficiently. NUTS employs a "No-U-Turn" criterion to determine the trajectory length during simulation. This criterion checks when the simulation turns back on itself, indicating that the exploration has reached an optimal point. These preliminary tuning steps facilitate establishing an equilibrium distribution for subsequent sampling. Chains that have successfully converged result in samples that accurately depict real-world data, thereby facilitating consistent and reliable forecasting during subsequent phases. Additionally, the number of iterations of initializer is set to $200,000$ to ensure a large number of initializations for finding a good starting point in the parameter space, which helps stabilize the Markov Chain Monte Carlo (MCMC) process, especially in complex models. We set the $random_seed=2000$ to ensure that results are reproducible by fixing the sequence of random number generation, providing consistency in outcomes across different runs.

During each MCMC step, a set of proposed parameters ($\Omega$) is required to drive the generation of a new time series reflecting newly infected cases. This series matches the observed data's length and is generated by utilizing the deterministic model introduced in (\ref{discrete_new_SEIR_eq}). The initiation point for the MCMC process follows the provision of an initial estimate of the posterior distribution of the model's parameters by the \textit{ADVI} approximation. Once the \textit{ADVI} approximation provides an initial estimate of the posterior distribution's parameters, the MCMC process begins. Throughout this process, the NUTS algorithm systematically draws and subsequently accepts or rejects the samples such that the resulting samples reflect the posterior distribution $p(\Omega | I_t) \sim p(I_t | \Omega) p(\Omega)$. In this context, $p(\Omega)$ is the prior distribution of the parameters (delineated in section \ref{sec:gen_algo} and $p(I_t | \Omega)$ represents the likelihood defined in \eqref{likelihood} for the real-world data given the parameters.  We chose the \textit{Poisson} distribution according to the model proposed in section \ref{sec:pseir_model}. 
\begin{equation}
    p(\hat{I_t} | \Omega) \sim Poisson(\mu = I_t)  
\label{likelihood}
\end{equation}
Likelihood quantifies how well the model, characterized by the model's parameters, explains or fits the observed data. Here, the likelihood is the product over local likelihoods quantifying the similarity between the model outcome for one data point ($\hat{I_t}$)  at time $t$ and the corresponding real-world data point $I_t$. Specifically, for each set of parameters acquired through sampling, the model equations outlined in \ref{discrete_new_SEIR_eq} come into play, driving the generation of data points for the new time series. The creation of these data points hinges on the proposed parameter values and the model's initial conditions. These newly generated time series unveil potential scenarios that seamlessly align with the observed data and the uncertainties encapsulated within the posterior distribution.


\bibliographystyle{Science}
\bibliography{scifile}

\end{document}